\documentclass[prb,twocolumn,preprintnumbers,amsmath,amssymb,floats,citeautoscript,nobalancelastpage,showpacs]{revtex4}

\usepackage{graphicx}
\usepackage{dcolumn}
\usepackage{bm}
\usepackage{times}

\begin{document}

\title{Low-temperature magnetotransport of narrow-gap semiconductor FeSb$_{2}$}

\author{H.~Takahashi}
\author{R.~Okazaki}
\author{Y.~Yasui}
\author{I.~Terasaki}

\affiliation{Department of Physics, Nagoya University, Nagoya 464-8602, Japan}

\begin{abstract}
We present a study of the magnetoresistance and Hall effect in the narrow-gap semiconductor FeSb$_{2}$ at low temperatures.
Both the electrical and Hall resistivities show unusual magnetic field dependence in the low-temperature range where a large Seebeck coefficient was observed.
By applying a two-carrier model, we find that the carrier concentration decreases from 1 down to 10$^{-4}$ ppm/unit cell and the mobility increases from 2000 to 28000 cm$^{2}$/Vs with decreasing temperature from 30 down to 4 K. 
At lower temperatures, the magnetoresistive behavior drastically changes and a negative magnetoresistance is observed at 3 K. 
These low-temperature behaviors are reminiscent of the low-temperature magnetotransport observed in doped semiconductors such as As-doped Ge, which is well described by a weak-localization picture.
We argue a detailed electronic structure in FeSb$_{2}$ inferred from our observations.     
\end{abstract}

\pacs{72.20.My, 72.20.Pa}

\maketitle

\section{Introduction}
Fe-based narrow-gap semiconductors have attracted considerable attention because of their unusual transport and magnetic behaviors, which closely resemble those of several rare-earth compounds termed ``Kondo insulators".
In 4$f$-electron Kondo insulators such as YbB$_{12}$, Ce$_{3}$Bi$_{4}$Pt$_{3}$, and CeNiSn,\cite{kondo1,kondo2,kondo3,kondo5,kondo6} a narrow gap is formed by the hybridization between localized 4$f$ and conduction electrons, leading to unique physical properties. 
The magnetic susceptibility obeys the Curie-Weiss law described by the local 4$f$ moment at high temperatures, followed by a broad maximum, and is suppressed at low temperatures. 
A large Seebeck coefficient due to a large density of states near the Fermi level was observed in the low-temperature range.
In analogy to the 4$f$-electron system, the narrow-gap semiconductor FeSi has been intensively studied as a prototype of the 3$d$-electron Kondo insulator.\cite{kondo1} 
This material displays an unusual crossover from high-temperature metal to low-temperature insulator,\cite{Kondo4,FeSi1} reminiscent of the 4$f$-electron Kondo insulators. 
In the photoemission experiments, however, the measured electronic structure has no distinct features relevant to a Kondo picture and is qualitatively explained within the band calculations by the density functional theory without many-body effects,\cite{FeSi2,FeSi3} puzzling the precise role of electron correlation in the formation of a narrow gap in this compound.

FeSb$_{2}$ is another candidate for the Fe-based Kondo insulator,\cite{FeSb1} the crystal structure of which is shown in the inset of Fig. 1. 
This material shows a unique temperature dependence of the magnetic susceptibility similar to that of FeSi.
The resistivity is well characterized by thermal activation with two energy gaps of $\Delta _{1}\sim$ 30 meV and $\Delta _{2}\sim$ 5 meV.\cite{FeSb2,FeSb5,FeSb9,FeSb10} The novelty of this compound is best highlighted by the colossal Seebeck coefficient $S$ $\simeq$ $-$45 mV/K near 10 K, where the smaller gap is open.\cite{FeSb2} Such a huge value of $S$ immediately indicates a novel mechanism, such as the strong electron correlation expected in Kondo insulators, which is pointed out by the thermodynamic and optical measurements.\cite{FeSb3, FeSb4, FeSb5} 
The Seebeck coefficient and resistivity, however, are extremely sensitive to the sample quality, and the reported maximum $S$ values span from $-$500 $\mu$V/K up to $-$45 mV/K.\cite{FeSb2, FeSb5, FeSb9, FeSb10, FeSb11}
A systematic study of the impurity effects on the transport properties of FeSb$_{2}$ revealed that the Seebeck coefficients are simply related to the carrier concentration doped by impurities, not to the electron correlations, as seen in conventional semiconductors.\cite{FeSb6}

In this paper, we present the magnetotransport study in FeSb$_{2}$ single crystals, especially for the investigation of its unusual electronic state at low temperatures. We conduct a two-carrier analysis for the conductivity tensor, which provides us with a proper evaluation of the mobility and carrier concentration.
The carrier concentration decreases from 1 down to 10$^{-4}$ ppm/unit cell with decreasing temperature from 30 down to 4 K. 
The small gap of 6 meV obtained from the temperature variation of the carrier concentration agrees well with the previous results.
The mobility reaches a large value of 28000 cm$^{2}$/Vs at 4 K, which enables us to observe the large Seebeck coefficients even in such a low carrier concentration in FeSb$_{2}$. At lower temperatures, the magnetoresistive behavior dramatically changes from positive to negative. The negative magnetoresistance is generally found in doped semiconductors at low temperatures and derives from a weak localization of carriers,\cite{fz2,fz4} implying that a similar electronic structure with a considerable impurity level is responsible for the unusual transport properties in FeSb$_{2}$.

\section{Experiment}
High-quality single crystals of FeSb$_{2}$ were grown by a self-flux method using metal powders of 99.999 $\%$ (5N) pure Fe and 99.9999 $\%$ (6N) pure Sb, as described in Ref.\:\onlinecite{FeSb6}.
The single-crystalline nature was checked by Bragg spots in the Laue pattern.
The resistivity was measured from 100 down to 2 K.
The magnetic field dependence of electrical and Hall resistivities were measured up to 70 kOe in the temperature range from 30 down to 3 K.
These transport properties were measured using a conventional four-probe dc method in the Quantum Design Physical Property Measurement System (Quantum Design, Inc.). 
For the magnetoresistance measurements, we employed the transverse configuration, i.e., with the current perpendicular to the magnetic field. We observed no significant direction dependence with respect to the crystalline axes.
For measurements of the Hall resistivity, the magnetic field was swept from $+$70 to $-$70 kOe, and the Hall voltage was determined as $V_{H}$=$\frac{1}{2}[V(+H)-V(-H)]$, where $V$ is the voltage across the Hall terminals.
The Seebeck coefficient was measured using a steady-state method.
\section{Results and Discussion}
Figure 1 shows the temperature dependence of the resistivity $\rho$ and the Seebeck coefficient $S$ below 100 K under zero magnetic field. The resistivity exhibits an insulating behavior, including two upturns separated by a plateau near 20 K, indicating the existence of two energy gaps in FeSb$_{2}$, as shown in previous reports.\cite{FeSb2,FeSb5,FeSb9,FeSb10} The Seebeck coefficient changes its sign around 30 K and a large value of $|S|\simeq1400 $ $\mu$V/K is observed near 20 K.  

\begin{figure}[t]
\begin{center}
\includegraphics[width=8cm]{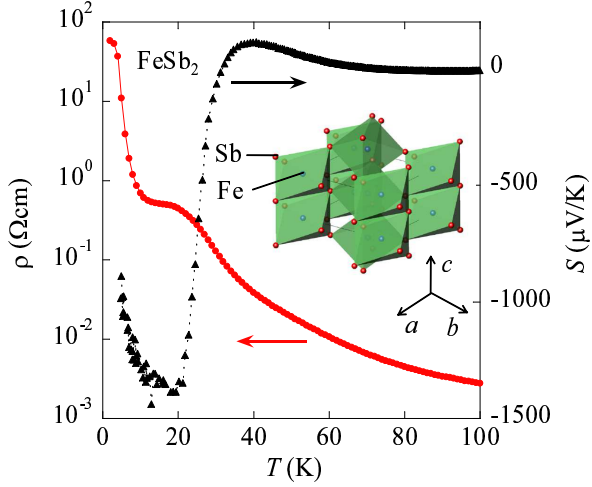}
\caption{(Color online) The temperature dependence of the resistivity $\rho$ (circle symbols, left axis) and the Seebeck coefficient $S$ (triangle symbols, right axis) under zero magnetic field. In the inset, the crystal structure of FeSb$_{2}$ is schematically shown.}
\end{center}
\end{figure}

In Figs. 2(a) and 2(b), we show the transverse magnetoresistance $\Delta  \rho/\rho \equiv [\rho_{xx}(H) -\rho_{xx}(0)]/\rho_{xx}(0)$, where $\rho_{xx}$ is the dc resistivity, measured at various temperatures from 3 to 30 K.
Above 15 K, a conventional quadratic magnetic field dependence of $\Delta  \rho/\rho$ is observed in low fields, but in high fields, it increases almost linearly with increasing magnetic field. At 8 and 10 K, $\Delta \rho/\rho$ tends to saturate in high fields and reaches 1.5 around 70 kOe. Such large values of $\Delta  \rho/\rho$ indicate high carrier mobility in this system. At 4 and 5 K, $\Delta  \rho/\rho$ is strongly suppressed at high magnetic fields. At 3 K, a negative magnetoresistance is observed.

\begin{figure}[t]
\begin{center}
\includegraphics[width=8cm]{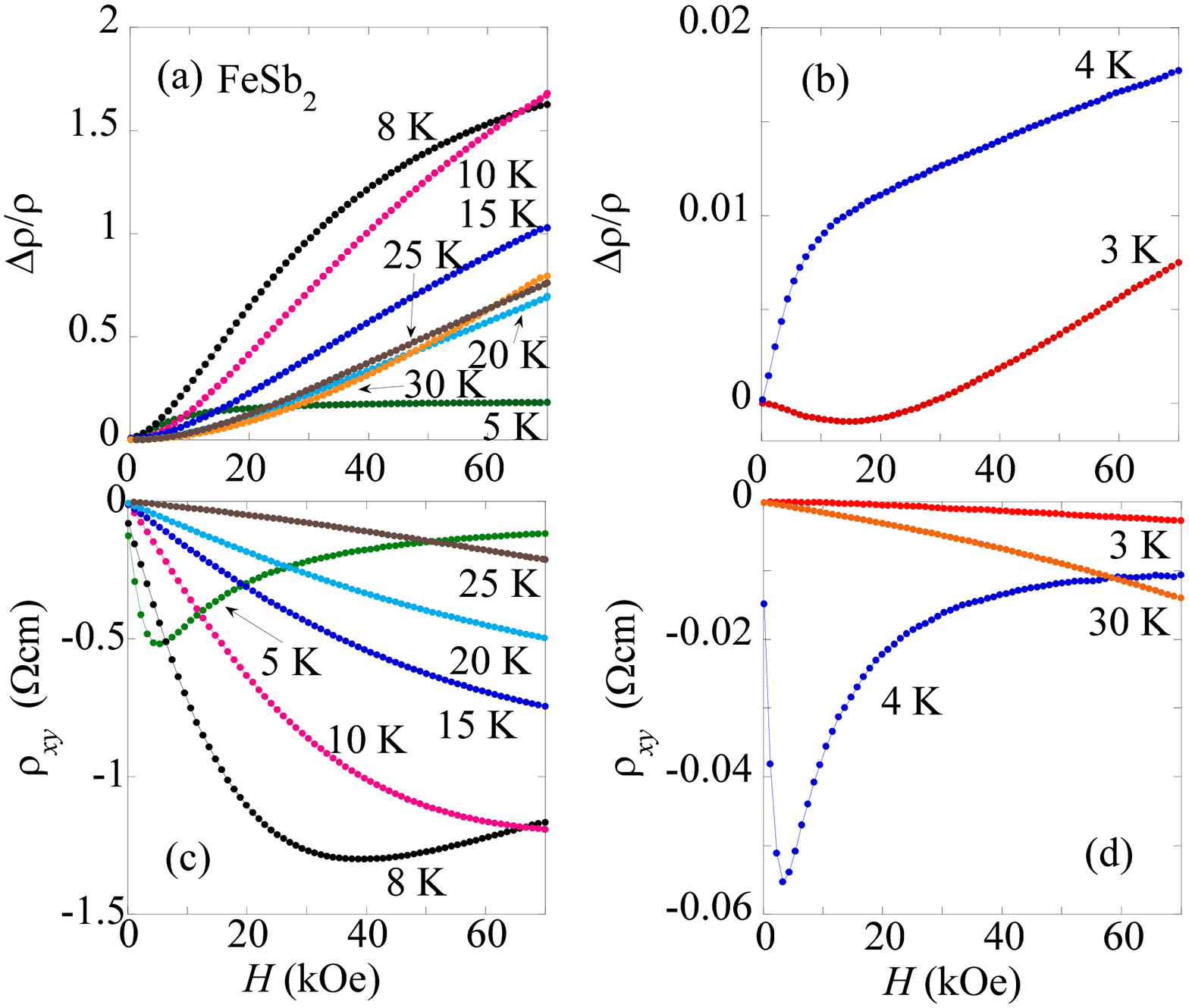}
\caption{(Color online) (a), (b) The transverse magnetoresistance $\Delta  \rho/\rho$, and (c), (d) the Hall resistivity $\rho_{xy}$, as a function of magnetic field measured at various temperatures. The vertical axes of (a) and (c) are nearly two orders of magnitude larger than those of (b) and (d), respectively.}
\end{center}
\end{figure}

The magnetic field dependence of the Hall resistivity $\rho_{xy}$ at several temperatures is shown in Figs. 2(c) and 2(d).
The Hall resistivity is highly nonlinear in $H$ from 4 to 20 K at the moderately high magnetic fields, in contrast to the typical $H$-linear dependence of $\rho_{xy}$ confirmed at 30 K.

\begin{figure}[t]
\begin{center}
\includegraphics[width=8cm]{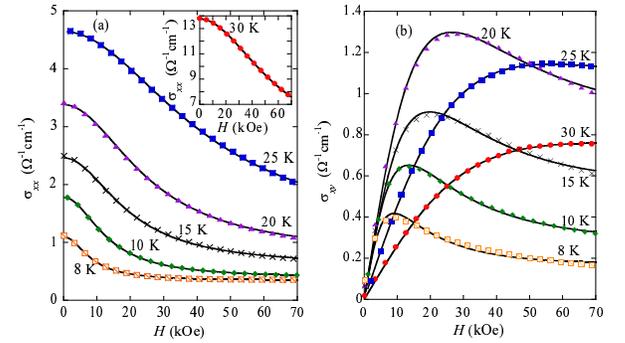}
\caption{(Color online) The magnetic field dependence of the conductivity tensor (a) $\sigma_{xx}$ and (b) $\sigma_{xy}$ from 8 to 30 K. Solid curves are the calculation using Eqs. (\ref{eq:sxx}) and (\ref{eq:sxy}).}
\end{center}
\end{figure}

\begin{figure}[t]
\begin{center}
\includegraphics[width=8cm]{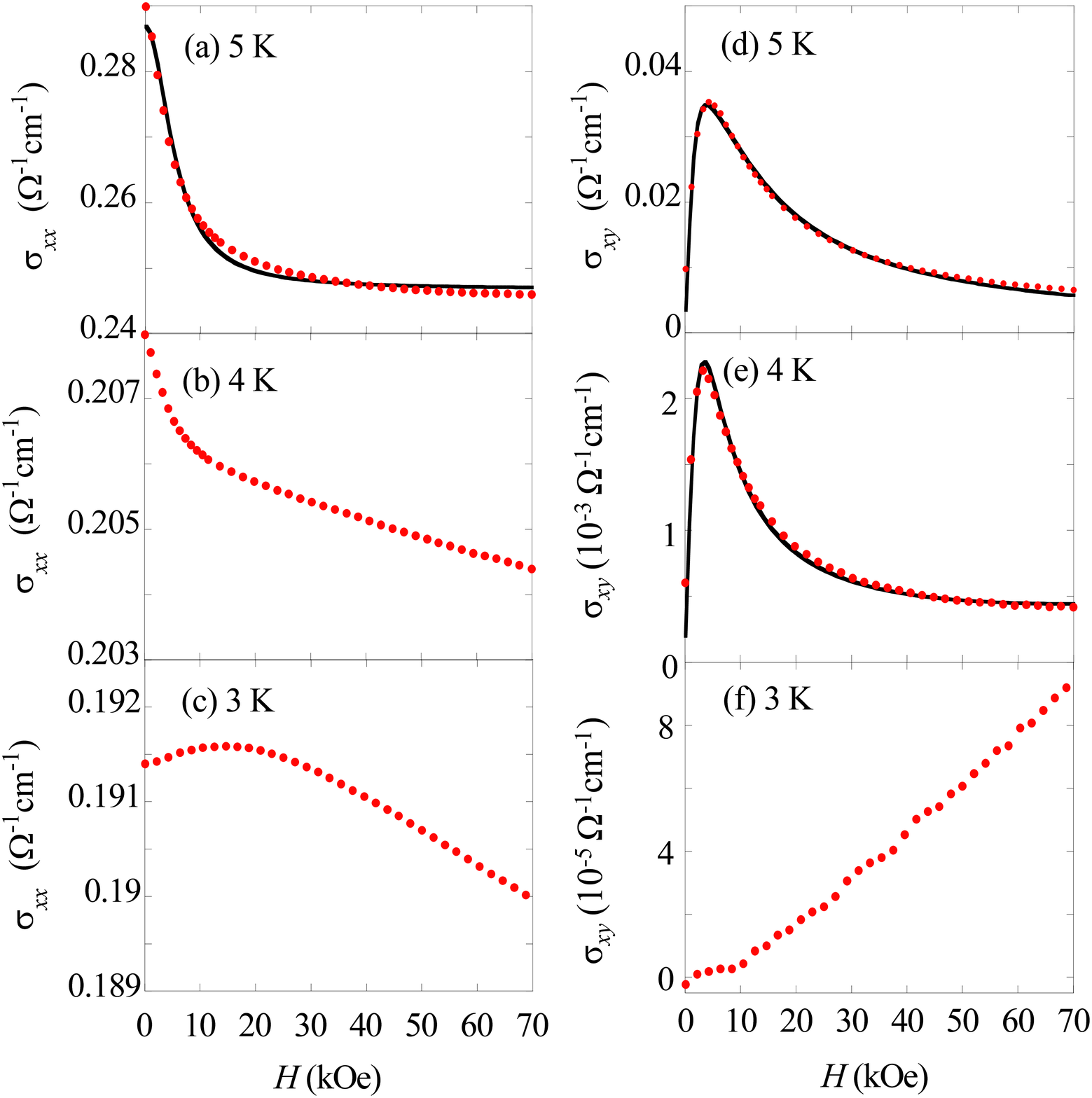}
\caption{(Color online) The magnetic field dependence of the conductivity tensor (a)--(c) $\sigma_{xx}$ and (d)--(f) $\sigma_{xy}$ from 3 to 5 K. Solid curves are the fitting calculation using Eqs. (\ref{eq:sxx}) and (\ref{eq:sxy}). $\sigma_{xx}$ below 4 K and $\sigma_{xy}$ at 3 K cannot be fitted with these equations.}
\end{center}
\end{figure}

To properly evaluate the carrier concentration $n$ and the mobility $\mu$, we now discuss the magnetic field dependence of the conductivity tensor.
In Figs. 3(a) and 3(b), we show $\sigma_{xx}=\rho_{xx}/(\rho_{xx}^{2}+\rho_{xy}^{2})$ and $\sigma_{xy}=-\rho_{xy}/(\rho_{xx}^{2}+\rho_{xy}^{2})$, respectively, in the temperature range between 8 and 30 K. The magnetic field dependence of $\sigma_{xx}$ and $\sigma_{xy}$ systematically changes with decreasing temperature. 
Figure 4 shows $\sigma_{xx}$ and $\sigma_{xy}$ below 5 K.
The magnetic field dependence of $\sigma_{xx}$ below 4 K is clearly different from that of the above temperatures, and a drastic change of $\sigma_{xy}$ curves is observed from 4 to 3 K. 
In the Boltzmann transport theory, the magnetic field dependence of the conductivity tensor is expressed by using $n$ and $\mu$. 
We here analyze $\sigma_{xx}$ and $\sigma_{xy}$ by using a two-carrier model in which one carrier is of high mobility and the other is of low mobility ($\mu H\ll 1$).
In this case, the $H$ dependence of the conductivity tensor is described as\cite{sigma}
\begin{equation}
\sigma_{xx} (H)=n_{xx}e\mu_{xx}\frac{1}{1+(\mu_{xx}H)^{2}}+C_{xx}, 
\label{eq:sxx}
\end{equation}
\begin{equation}
\sigma_{xy} (H)=n_{xy}e\mu_{xy}^{2}H\left[\frac{1}{1+(\mu_{xy}H)^{2}}+C_{xy}\right].
\label{eq:sxy}
\end{equation}
Here $n_{xx}$ ($n_{xy}$), $\mu_{xx}$ ($\mu_{xy}$), and $C_{xx}$ ($C_{xy}$) are the carrier concentrations, the carrier mobilities, and the low-mobility components for $\sigma_{xx} $ ($\sigma_{xy}$), respectively.
The best agreements of $\sigma_{xx}$ and $\sigma_{xy}$  with Eqs. (\ref{eq:sxx}) and (\ref{eq:sxy}), respectively, are shown by solid curves in Figs. 3 and 4. 
However, the data of $\sigma_{xx}$ at 3 and 4 K and those of $\sigma_{xy}$ at 3 K are unadaptable to the above two-carrier model, and we will discuss this reason later.

\begin{figure}[t]
\begin{center}
\includegraphics[width=8cm]{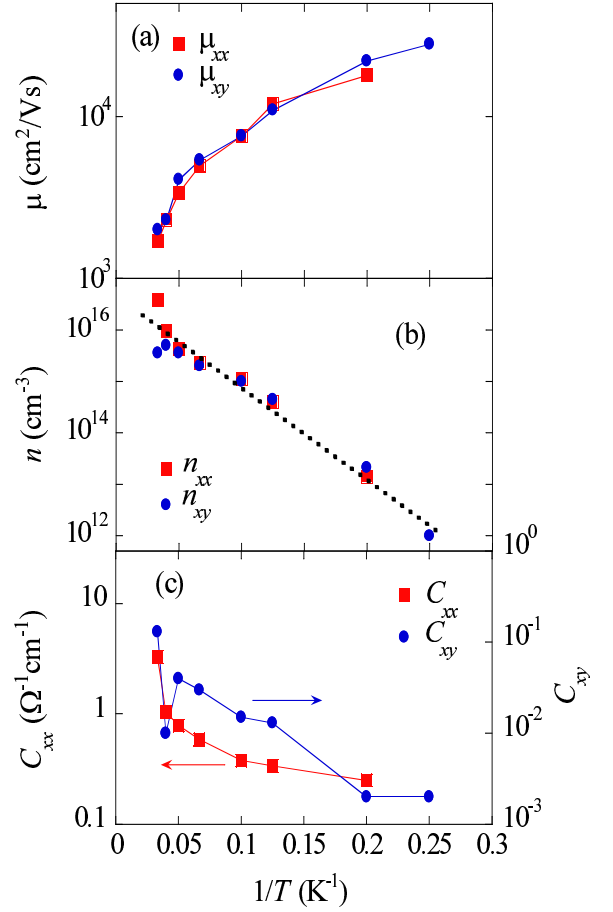}
\caption{(Color online) 1/$T$ dependence of (a) the carrier mobility $\mu$ and (b) the carrier concentration $n$, and (c) the low-mobility components $C_{xx}$ and $C_{xy}$, obtained from the fitting results for the conductivity tensor. The dotted line (b) is the fitting result by the thermal activation function $n\propto\exp(-\Delta  /2k_{B}T)$, which yields $\Delta   = 6$ meV.}
\label{fig:uxxuxy}
\end{center}
\end{figure}

In Figs. 5(a) and 5(b), we show the 1/$T$ dependence of the carrier mobility and the carrier concentration obtained from the fitting for the conductivity tensor, respectively. 
The carrier mobilities $\mu_{xx}$ and $\mu_{xy}$, which are obtained independently from $\sigma_{xx}$ and $\sigma_{xy}$, are in excellent agreement with each other, justifying our evaluation for these quantities within the two-carrier model.
This is also evidenced by the quantitative agreement between $n_{xx}$ and $n_{xy}$.
The low-mobility components $C_{xx}$ and $C_{xy}$ from the fitting results are displayed in Fig. 5(c).
The mobility $\mu_{\rm low}$ estimated from $C_{xx}$ and $C_{xy}$ is about 25 \% of the high-mobility components at 10 K.
Thus, $\mu_{\rm low} H \ll 1$ holds at low fields and even at high fields, and our fitting model using Eqs. (1) and (2) is still valid because the contributions of the low-mobility components to $\sigma_{xx}$ and $\sigma_{xy}$ are much smaller than those of high-mobility carriers.

Let us first discuss the carrier mobility in this material.
As shown in Fig. 5(a), the mobility of the dominant carriers is about 2000 cm$^{2}$/Vs at 30 K and increases with decreasing temperature. At 4 K, the mobility reaches a significantly large value of $\mu\simeq 28000$ cm$^{2}$/Vs, comparable to that of high-purity Si or Ge semiconductors.\cite{Si2,Ge2} 
Such a large value of the mobility at low temperatures has been also observed in several Kondo insulators and charge-density-wave systems,\cite{CeNiSn, MoO} because of an enhancement of the scattering time due to gap opening.
Previous studies of the transport properties have shown a large mobility in FeSb$_{2}$.\cite{FeSb10,FeSb7}   
The Hall mobility $\mu_{H}=|R_{H}|/\rho$, which is obtained from the Hall coefficient $R_{H}$ and the resistivity $\rho$,\cite{FeSb10} is different from the present results: $\mu_{H}$ exhibits a broad peak around 8 K and its maximum value is five times smaller than ours. This deviation from our results is clearly due to the evaluation method in which only a single band is assumed. 
Hu $et$ $al$. reported a giant carrier mobility of $\mu\simeq 8\times10^{4}$ cm$^{2}$/Vs near 10 K, which was estimated only from the magnetic field dependence of the Hall coefficient within a two-carrier model.\cite{FeSb7} Their raw data of $\rho_{xy} (H)$ at 10 K is, however, much smaller than ours at the same temperature and can be plotted into our data between 30 and 25 K, which implies that $\mu$ is of the order of 10$^{3}$ cm$^{2}$/Vs. 
The fitting model used in Ref.\:\onlinecite{FeSb7} to evaluate both the carrier concentration and the mobility is only valid for very-high-mobility electron systems in a nonquantizing low-field range, in which $\rho_{xy}(H)$ exhibits a strong nonlinear $H$ dependence,\cite{model} incompatible with the nearly $H$-linear dependence of $\rho_{xy} (H)$ observed in Ref.\:\onlinecite{FeSb7}.
Moreover, in the high-field range, the model equation of the Hall coefficient does not depend on $H$, and the values of the mobility and carrier concentration cannot be extracted independently using this model.\cite{model} 
Thus, the model might greatly overestimate the carrier mobility in this system.

We next discuss the carrier concentration shown in Fig. 5(b). 
With decreasing temperature from 30 down to 4 K, it rapidly decreases from 10$^{16}$ cm$^{-3}$ (1 ppm/unit cell) to 10$^{12}$ cm$^{-3}$ (10$^{-4}$ ppm/unit cell). 
As shown by the dotted line, it is well fitted by the thermal activation function $n\propto\exp(-\Delta  /2k_{B}T)$, which yields $\Delta=6$ meV.
The systematic study of the impurity effects on the transport properties indicates that the electric conduction in this temperature range is governed by extrinsic carriers; the main carriers are excited from the impurity level, which locates slightly below the bottom of the conduction band.\cite{FeSb6} 
In this picture, $\Delta $ corresponds to the energy gap between the impurity level and the bottom of the conduction band.
In other words, the gap $\Delta$ is unlikely to be a gap in the Kondo insulators.

Below 4 K,  we could not fit $\sigma_{xx}$ using Eq. (\ref{eq:sxx}).
As seen in Fig. 1(b), the magnitude of $\Delta  \rho/\rho$ is extremely small compared to those above 5 K, and the negative magnetoresistance is observed around low magnetic fields at 3 K.
These behaviors have been widely observed in doped semiconductors in the low-temperature range because of a weak electronic localization.\cite{fz2,fz4}
In the case of As-doped Ge, impurity carriers show an electric conduction at very low temperatures, 
where the thermal excitation into the conduction band is negligibly small. 
The carriers are weakly localized by a random potential, leading to a negative magnetoresistance.\cite{fz4}
The striking similarity of low-temperature magnetoresistive behaviors between FeSb$_{2}$ and conventional semiconductors suggests a considerable impurity level in the electronic structure of FeSb$_{2}$: The impurity level exists slightly below ($\sim$ 6 meV) the bottom of the conduction band and carriers in the impurity level are predominant on the  electrical transport at 3 K.
   
Finally, we discuss the Seebeck coefficient $S$ with a calculation using a nondegenerate model expressed as \cite{nonde}
\begin{equation}
S= \pm \frac{k_{B}}{e}(\frac{\Delta  }{k_{B}T}+\delta+\frac{5}{2}) ,
\label{eq:s}
\end{equation}
where $\delta$ is a scattering parameter.
Now we calculate the value of $S$ using $\Delta=6$ meV, which is estimated from the 1/$T$ dependence of the carrier concentration, and obtain $|S| \sim 600$ $\mu$V/K at 20 K.
Here we assume $\delta$ = 3/2, which is used in conventional semiconductors.
The calculated value is close to, but nearly two times smaller than, the experimental results of $S=-800\sim-1400$ $\mu$V/K in Refs. \:\onlinecite{FeSb6} and \:\onlinecite{FeSb8}.
An electron-electron correlation is unlikely to be the origin of this deviation because it has an insignificant effect on the Seebeck coefficient in an insulating regime.\cite{FeSb12}
On the other hand, the phonon-drag effect may play an essential role for an enhancement of the Seebeck coefficient in the low-temperature region, as seen in many pure semiconductors.\cite{InSb, Ge, Si}
The phonon-drag term in the Seebeck coefficient can be approximately expressed as 
$S_{p}=\beta v_{p} l_{p}/\mu T$,
where $v_{p}$ and $l_{p}$ are the velocity and the mean free path of a phonon, respectively, and $0<\beta<1$ is a parameter which characterizes the relative strength of the electron-phonon interaction.\cite{Si, ph}
From the high thermal conductivity below 30 K, a long mean free path of $l_{p}\simeq10 $ $\mu$m at 20 K was estimated by Bentien $et$ $al$.\cite{FeSb2, FeSb5, FeSb10} 
In addition, they calculated the sound velocity to be 3000 m/s.
Using these parameters, $S_{p}$ at 20 K is estimated to be nearly 3 mV/K at maximum, which suggests that the phonon-drag effect can yield a measurable contribution to the Seebeck coefficient in this material.
On the other hand, it is still difficult to understand the colossal Seebeck coefficient ($S\sim-$45 mV/K at 10 K) that was observed in Ref. \:\onlinecite{FeSb2} within the above description.

\section{conclusion}
We have measured the magnetic field dependence of the electrical and Hall resistivity of pure FeSb$_{2}$ single crystals and analyzed the conductivity tensor by using a two-carrier model to evaluate the carrier concentration and mobility.
The carrier concentration decreases from 1 down to 10$^{-4}$ ppm/unit cell with decreasing temperature from 30 down to 4 K. The mobility significantly increases with decreasing temperature and reaches 28000 cm$^{2}$/Vs at 4 K.
At lower temperatures, the transport behavior changes drastically and a negative magnetoresistance is observed. 
In doped semiconductors, the negative magnetoresistance due to a weak localization is generally found at low temperatures, implying a considerable impurity level in the electronic structure of FeSb$_{2}$. 
The temperature variation of the carrier concentration indicates the existence of an energy gap ($\sim$ 6 meV) between the bottom of the conduction band and the impurity level.
The magnetotransport reported in this paper strongly suggests that the low-temperature transport of FeSb$_{2}$ is well understood as an extrinsic semiconductor with ppm-level impurity. The phonon-drag effect may work, but the electron correlation effect is secondary.

\section*{Acknowledgements}
The authors would like to thank M. Sato for the initial motivation and collaboration at an early stage of this work. The authors also appreciate H. Kontani and A. Kobayashi for their helpful discussion. They are also indebted to Q. Li for sharing his unpublished data. This work was partially supported by the Strategic Japanese-Finland Cooperative Program on  ``Functional Materials," JST, Japan.

\end{document}